# Measuring, in solution, multiple-fluorophore labeling by combining Fluorescence Correlation Spectroscopy and photobleaching

**Antoine Delon[a]\*, Irène Wang[a], Emeline Lambert[b],[c], Silva Mache[b], Régis Mache[b], Jacques Derouard[a], Vincent Motto-Ros[a],[d], Rémi Galland[a]**

[a] Univ. Grenoble I / CNRS, Laboratoire de Spectrométrie Physique UMR 5588, BP 87, 38402 Saint Martin d'Hères

[b] Univ. Grenoble I / CNRS, Laboratoire de Physiologie Cellulaire et Végétale UMR 5168, BP 53, 38041 Saint Martin d'Hères

[c]Present adress: Univ. Grenoble I / IAB, Equipe13, 38706 La Tronche Cedex

[d]Present adress: Univ. Lyon I / CNRS, LaSIM UMR 5579, 69622 Villeurbanne Cedex

\* corresponding author : adelon@ujf-grenoble.fr

# Abstract

Determining the number of fluorescent entities that are coupled to a given molecule (DNA, protein, *etc*.) is a key point of numerous biological studies, especially those based on a single molecule approach. Reliable methods are important, in this context, not only to characterize the labeling process, but also to quantify interactions, for instance within molecular complexes. We combined Fluorescence Correlation Spectroscopy (FCS) and photobleaching experiments to measure the effective number of molecules and the molecular brightness as a function of the total fluorescence count rate on solutions of cDNA (containing a few percent of C bases labeled with Alexa Fluor 647). Here, photobleaching is used as a control parameter to vary the experimental outputs (brightness and number of molecules). Assuming a Poissonian distribution of the number of fluorescent labels per cDNA, the FCS-photobleaching data could be easily fit to yield the mean number of fluorescent labels per cDNA strand ($\cong 2$). This number could not be determined solely on the basis of the cDNA brightness, because of both the statistical distribution of the number of fluorescent labels and their unknown brightness when incorporated in cDNA. The statistical distribution of the number of fluorophores labeling cDNA was confirmed by analyzing the photon count distribution (with the cumulant method), which showed clearly that the brightness of cDNA strands varies from one molecule to the other. We also performed complementary continuous photobleaching experiments and found that the photobleaching decay rate of Alexa Fluor 647 in the excited state decreases by about 30% when incorporated into cDNA, while its non radiative decay rate is increased such that the brightness of individual Alexa labels is decreased by 25% compared to free Alexa dyes.

# 1. Introduction

The multiple labeling of biological molecules, such as proteins or DNA strands, with fluorophores, is a key point for molecular imaging in microscopy, the investigation of biomolecular processes,[1, 2] biochips, DNA sequencing,[3, 4, 5, 6, 7] *etc…* The knowledge of the number of fluorophores incorporated into the molecules is an important issue since it determines the brightness of the molecules to be detected and affects the detection limit. This is especially true for single molecule measurements,[8] that have, for instance, revealed aggregation that would have not been detected otherwise.[9] When the goal is to measure a single molecule, it is important to be able to distinguish aggregates of various sizes, which implies a proper understanding of multiple labeling. Especially demanding are experiments that aim at quantifying aggregation of molecular systems.[10, 11] One possible method to measure the number of dyes present in an aggregate is to count the number of steps observed in the time trace of the fluorescence of single molecules before they are completely photobleached.[1, 12, 13] More sophisticated methods have also been proposed, that rely on a statistical analysis of the fluorescence trajectories during photobleaching or on the deconvolution of the intensity distribution of single molecules.[14, 15] However, all of them require the molecules to be immobilized on a substrate and observed using a high detection efficiency and specificity. Moreover a large number of cases need to be analyzed to get the whole statistical distribution of labeling and brightness. This may not be compatible with any biological system. Therefore bulk experiments, that would nevertheless reveal information at a single molecule level, would be of high interest. This is in exactly the case of methods based on the analysis of intensity fluctuations, such as the Photon Counting Histogram (PCH), or the Fluorescence Intensity Distribution Analysis (FIDA).[16-20] Although these are not, strictly speaking, single molecule techniques, they exploit the statistical properties of the fluorescence fluctuations from a limited amount of molecules present in the observation volume at a given time. Note that these methods often take benefit from the possibility to titrate the sample in order to vary the degree of aggregation, which is not always doable.

Here we present a new approach to estimate the number of fluorescent labels by combining Fluorescence Correlation Spectroscopy (FCS) and photobleaching. We have studied single strand cDNA labeled with Alexa-Fluor-647 as a typical system of molecules randomly labeled with a small number of fluorophores. Although the measurements are primary averaged over an ensemble of molecules, we will show that it is still possible to extract information concerning the statistical distribution of labeling. Our method relies on the measurement of the effective brightness of cDNA molecules in solution, defined as the ratio of the total fluorescence signal to the effective mean number of molecules in the observation volume and on the analysis of its behavior, while the sample is being sequentially photobleached. Photobleaching is thus used as an additional degree of freedom (like titration) that makes it possible to extract more information from the system. The experiments have been carried out using two main tools: i) FCS to measure the effective mean number of molecules in the observation volume,[21] together with their brightness and ii) photobleaching of the molecules confined in small volumes (with dimensions a few tens of μm). In addition, the fluctuations of intensity have been studied, using the Fluorescence Cumulant Analysis (FCA), to confirm the statistical nature of the number of fluorophores labeling the cDNA strands.[22] The FCA approach has the other advantage of offering a powerful theoretical framework to interpret the brightness measurements. The data are then analysed and interpreted in terms of multiple labeling. Finally, our results give some insights into the photophysical properties (non radiative and photobleaching decay rates) of Alexa fluorophores when grafted to the cDNA backbone as compared to free fluorophores.

The material and experimental set-up are presented in Section 2; Section 3 introduces the theoretical framework; Section 4 is devoted to the experimental results (continuous photobleaching, FCS and sequential photobleaching) and their discussion; the conclusion is made in Section 5.

## 2. Material and methods

### 2.1. Samples and preparation

The experiments have been carried out on two types of single strand cDNA of different size:

i) The so-called long cDNA was prepared from RNA obtained by *in vitro* transcription of the *EcoRI* linearised plasmid pTZ19/PLS-Ta using T7 RNA polymerase (MaxiScript SP6/T7 *in vitro* transcription kit, Ambion).[23] The produced RNA was reverse transcribed in the presence of 500 µM dATP, dGTP and dTTP. For dCTP three different solutions were used. The final concentration of dCTP was kept at 30 µM but contained either 67% (20 µM) or 100% Alexa-Fluor-647 labeled dCTP (Invitrogen).

ii) To produce the short cDNA, part of the pTZ19/PLS-Ta sequence was cloned after PCR amplification (primers 5'-CATTAAGGCCTAATTTATGTCG-3' and 5'-AAGCCCGCACTGTCAGG-3') into the vector pCR2.1 (Invitrogen). The resulting plasmid was transcribed with T7 RNA polymerase after linearization with *HindIII* and the obtained RNA was reverse transcribed under the same conditions as for the long cDNA (see above). The sequence of the primer used for cDNA synthesis was 5'-CGTGCCTGTTCTTCGCGTCC-3' for the long and the short cDNA.

Unincorporated nucleotides were separated from the cDNAs by migration in a 1.5 % agarose gel followed by elution of the cDNAs from the agarose gel (High Pure PCR Product Purification Kit, Roche). The long cDNA consists of 306 bases including 74 dCTP deoxynucleotides in its sequence and the short cDNA consists of 147 bases including 45 dCTPs in its sequence.

The presence of Alexa-Fluor-647 labeled deoxynucleotides in the reverse transcription reaction diminishes the quantity of the produced cDNA considerably. No cDNA is detectable if only Alexa-Fluor-647 labeled dCTP is present in the reaction. In the following, all cDNA synthesis was performed in the presence of 67% Alexa-Fluor-647-dCTP.

A specific problem that arose with the short cDNA concerned dimerisation: the short cDNA migrates at a size that is larger than 200 nucleotides. This band corresponds to cDNA dimers. The two strands can be separated by denaturation of the sample at 95°C for 5 mn before loading. However, in the following, we will only consider the case of native short cDNA strands (*i.e.* hybridized), unless specified.

Free Alexa-Fluor-647 (Alexa) and Cy5 dyes were purchased from Invitrogen Molecular Probes and stock solution were prepared without further purification.

## 2.2 Photobleaching

In order to achieve photobleaching of the fluorophores in solution in a reasonable amount of time (a few minutes), with a reasonable laser power (a few mW), the solutions were confined in small volumes. Sample solutions were thus confined into poly(dimethylsiloxane) (PDMS) micro-wells. These micropatterned surfaces were prepared in two steps. First the negative pattern was obtained by standard UV lithography on Si substrate. A regular array containing pillars of 60 μm in diameter and 50μm in height with a pitch of 350 μm was fabricated on the Si wafer. Then, a 10:1 mixture of PDMS base with curing agent (Sylgard 184, Dow Corning) was poured directly on the wafer and baked at 65°C for 2 hours. After peeling off the mold, the 2-mm thick PDMS micropatterned slabs were treated with atmospheric RF plasma to make the surface hydrophilic and facilitate PDMS bonding with glass slides. After this treatment, a drop of the solution of interest (Alexa or cDNA in water) was spread on the surface. Excess solution was removed and a glass coverslip was placed on top of the PDMS surface to cover the micro-wells. We controlled that the wells were filled with solution using an optical microscope in transmission. The experiments were performed within *c.a.* 3 hours following the sample preparation, in order to avoid evaporation.

The experimental setup consisted of a helium-neon laser for excitation (HRP170, Thorlabs) emitting 17 mW maximum power at 633 nm wavelength. This laser beam was sent through a beam expanding telescope and a focusing lens, then into an inverted microscope. In the object plane of the microscope objective (Olympus UMPlanFl, 10×), the laser spot had a Gaussian shape with a 230-µm waist (at $1/e^2$ radius) as measured using a uniformly fluorescent polymer film. This beam was large compared to the micro-wells (diameter 60 µm), so that the excitation intensity was supposed to be constant over a micro-well placed at the center of the laser spot. The excitation intensity was varied from 5 W/cm$^2$ to 18 W/cm$^2$. The fluorescence emission from the solution trapped in the micro-well was collected in an epifluorescence configuration by the microscope objective and imaged on a cooled CCD camera (Princeton Instruments). The fluorescence emission signal was obtained by averaging the camera pixels values in a central 26 µm square region. At each excitation power, the fluorescence decays obtained from three micro-wells were averaged.

Continuous photobleaching experiments were performed with aqueous solutions of Alexa, Alexa-dCTP, and Alexa labeled cDNA strands of both lengths. We checked that the so called photobleaching process is really irreversible, on the time scale of our experiment. The fluorescence signal followed a perfectly mono-exponential decay, for all the samples and excitation powers studied, as exemplified in Figure 1. The decay rate, $1/\tau_B$, depends linearly on the incident power for all samples and is related to the photobleaching cross-section, $\sigma_B$, through:

$$\frac{1}{\tau_B} = \sigma_B I_{ph} = \sigma_B \frac{2P}{\pi w^2} \qquad (1)$$

where $\tau_B$ is the time constant, $I_{ph}$ is the incident photon intensity (in photon/s/m$^2$), supposed uniform over the sample, $w$ is the waist of the excitation laser in the sample plane and $P$ is the total power of the excitation laser (in photon/s).

For the Alexa dye solution, the exponential decay confirms that our measurements are not distorted by other effects such as diffusion inside the sample or adsorption on the walls of the wells. We checked our measurements by determining the photobleaching quantum yield (*i.e.* the ratio between the photobleaching and absorption cross-sections). The absorption cross-section was obtained independently from transmission measurements on concentrated solutions. We found a photobleaching quantum yield of 2.4×10$^{-6}$ for Alexa and of 4.7×10$^{-6}$ for Cy5. The latter value is in good agreement with the value of 5×10$^{-6}$ determined elsewhere[24]. These results indicate that Alexa is approximately twice more photostable than Cy5, which is also consistent with the literature.[25]

For the cDNA strands, the mono-exponential behaviour means that, due to the weak degree of labeling, the Alexa labels of the same strand behave like independent fluorophores with negligible interactions between each other, though they may interact with the cDNA backbone. As a matter of fact, if self-quenching (due to Förster-type energy homo-transfer) would occur,[26, 27] the excited state lifetime of individual fluorescent labels (hence their brightness and the photobleaching decay rate) would vary during photobleaching, since the labeling density varies.

## 2.3 Fluorescence Correlation Spectroscopy

We have used a home made, though rather standard, experimental set-up to record the time trace of the fluorescence fluctuation intensity of the molecules in solution from which we could extract the autocorrelation function as mentioned in the theoretical section. Our set-up is mostly composed of an inverted microscope (Olympus IX70) working in the confocal configuration. The fibered incident laser beam of power 50 to 100 µW (637 nm, LDM635, Thorlabs), was firstly collimated and expanded (×2.4) through a telescope, then sent, *via* a periscope, to a dichroic mirror (z633rdc, Chroma), before being focused within the sample, through the water objective (60×, NA 1.2) of the inverted microscope. The fluorescence light was then spectrally filtered (HQ700/75m, Chroma) and focused on a multimode fiber (core diameter 100 µm), used as point detector. Note that a pair of lenses, located after the tube lens, produces an additional 3× magnification, such that the total magnification was about 180×. Finally, an Avalanche Photodiode, (SPCM-AQR-13, PerkinElmer) was connected to a home made data acquisition system and correlator. Typically a series of 5 acquisitions of 10 to 20 s was performed for each FCS measurement, to properly calculate the averaged autocorrelation curve and standard error of the mean. Data plotting and non linear least square fitting were performed using Origin software (OriginLab Corporation), to extract the effective number of molecules, $N_{eff}$, and effective brightness, $\varepsilon_{eff}$, as explained in the following section.

This set-up has also been used to sequentially photobleach the cDNA solutions, confined into the PDMS micro-wells, as previously described, in combination with FCS measurements. The corresponding protocol consisted in increasing the laser power to several mW, while monitoring the total fluorescence decrease, until it has typically dropped by a factor 1.5. Then the laser power was adjusted back to a power of a few tens of µW, to avoid photobleaching and saturation during the consecutive FCS measurement. When finished, a new photobleaching sequence begins and so on.

## 3. Theory

### 3.1 Fluorescence Correlation Spectroscopy

The Fluorescence Correlation Spectroscopy (FCS) is an experimental method that aims at measuring concentration (or number density), diffusion constant and brightness of molecules (or particles) in solution, cellular media, membranes, *etc*. This method exploits the temporal fluctuations of the fluorescence intensity emitted by a small amount of fluorescent molecules diffusing in and out a detection volume, generally under confocal geometry. Using single photon detector like avalanche photodiode, it is possible to detect and measure concentrations from about 0.1 nM to 1 µM. It is thus perfectly adapted to the present situation where we want to analyze cDNA solutions in the nM range, mainly to get information in terms of number concentration and molecular brightness.

For quantitative evaluation, the FCS analysis is performed using the so-called autocorrelation function $G(\tau)$, the amplitude of which is inversely proportional to the effective mean number of molecules in the observation volume, $N_{eff}$:

$$G(\tau) = \frac{\langle I(t)I(t+\tau)\rangle}{\langle I(t)\rangle^2} = 1 + \frac{\gamma_2}{N_{eff}} g(\tau) \qquad (2)$$

where $\gamma_2$ is a geometric factor that takes into account the exact shape of the observation volume. Using factorial cumulant analysis, we estimated this factor to be 0.23 (see next Section 3.2.2).

The time dependant part, $g(\tau)$, is characterized by the diffusion time, $\tau_D$, that is dependant upon the diffusion constant, $D$, and the radius of the confocal observation volume, $w_0$, according to:

$$\tau_D = w_0^2/4D \qquad (3)$$

Another important output of FCS experiments, which will be used throughout this paper, is the effective photon count rate per molecule, defined as the total photon count rate *CR* divided by the effective number of molecules:

$$CRM_{eff} = CR/N_{eff} \qquad (4)$$

For more details about FCS data analysis see for instance[21, 28]. Throughout this paper we will also refer to the so called effective brightness, $\varepsilon_{eff}$, defined as the number of photon counts per molecule, during a given binning time, $\delta t$ (that must be much smaller than the diffusion time):

$$\varepsilon_{eff} = CRM_{eff} \times \delta t \qquad (5)$$

The theoretical interest of the brightness will be introduced in the next section. Note that the subscript $_{eff}$ used for the number of molecules, the count rate per molecule or the brightness, refers to the fact that these quantities can be measured on a non uniform population of molecules, with a statistical distribution of brightness. In general, $\varepsilon_{eff}$ is different from the average brightness $\bar{\varepsilon}$, so is $N_{eff}$ from the actual number of molecules.

## 3.2 Fluorescence cumulant analysis

### 3.2.1 General properties

Like other methods developed to analyze intensity fluctuations,[17-20] Fluorescence Cumulant Analysis (FCA) provide information about the brightness distribution. However, contrary to the Photon Counting Histogram (PCH) or to the Fluorescence Intensity Distribution Analysis (FIDA), FCA is considerably simpler to implement and mathematically well suited to the present situation. The factorial cumulants, $\kappa_{[i]}$, are statistical quantities linearly related to the moments of the photon counts detected during a given binning time, $\delta t$.[29] Following Müller,[22] if the photon counts are those of a population of fluorescent molecules (in our case, cDNA strands) having a distribution of brightness characterized by its moments $\overline{\varepsilon^r}$, the factorial cumulant of order $r$ is simply given by (for a binning time, $\delta t$, much smaller than the diffusion time):

$$\kappa_{[r]} = \gamma_r N_{DNA} \overline{\varepsilon^r} \qquad (6)$$

where $\gamma_r$ is a geometric factor depending upon the shape of the observation volume and $N_{DNA}$ stands for the number of all cDNA molecules in the observation volume (*i.e.* including those having a nil brightness). Note that, correspondingly, the moments $\overline{\varepsilon^r}$ are also defined over the whole population of molecules. In case of one photon excitation, Huang *et al.* have shown that $\gamma_r$ is well approximated by the formulae $\gamma_r = 1/[(1 + F)r^{3/2}]$, except for $\gamma_1$, which is equal to 1.[30] $F$ is a parameter that accounts for the deviation of the observation volume from a perfect Gaussian shape.

### 3.2.2 Single valued brightness and low order cumulants

If the photon counts come from a homogenous solution of *N* fluorescent molecules in the observation volume, with all the same brightness, $\varepsilon$, one can calculate (using eq 6) the number of molecules, *N*, with the ratio:

$$N = \gamma_2 \frac{\kappa_{[1]}^2}{\kappa_{[2]}} \qquad (7a)$$

and the brightness, $\varepsilon$, from:

$$\varepsilon = \frac{1}{\gamma_2} \frac{\kappa_{[2]}}{\kappa_{[1]}} \qquad (7b)$$

Of course, in this case, $\varepsilon = \varepsilon_{eff} = \bar{\varepsilon}$ and $N = N_{eff}$. Here, we emphasize the fact that the number of molecules *N* and the brightness $\varepsilon$ calculated from the two first cumulants are strictly equivalent to those obtained by FCS, providing the binning time, $\delta t$, is properly chosen.

Note that the *F* parameter, which characterizes our setup, can be determined using measurements on homogeneous dye, since the ratio $\kappa_{[2]}^2/\kappa_{[1]}\kappa_{[3]}$ is equal to $\gamma_2^2/\gamma_3 = 3^{3/2}/[8(1 + F)]$ in this case. Using a pure solution of Alexa, we found $F = 0.56$, that is $\gamma_2 = 0.23$, instead of 0.35 for a perfect Gaussian shape.

### 3.2.3 Brightness distribution and low order cumulants

If the solution of interest is a mixture of molecules (*i.e.* cDNA strands) having a distribution of brightness, characterized by a first and a second moment, $\bar{\varepsilon}$ and $\overline{\varepsilon^2}$, the two first cumulants read:

$$\kappa_{[1]} = N_{DNA}\,\bar{\varepsilon} \tag{8a}$$

$$\kappa_{[2]} = \gamma_2 N_{DNA}\,\overline{\varepsilon^2} \tag{8b}$$

Let us assume that the brightness distribution of cDNA molecules is only due to the distribution of the number, $n$, of Alexa fluorophores attached to a cDNA strand. If $\varepsilon_A$ is the brightness of individual fluorophores, we thus have $\bar{\varepsilon} = \varepsilon_A \bar{n}$ and $\overline{\varepsilon^2} = \varepsilon_A^2 \overline{n^2}$. For a Poissonian distribution of average value $\bar{n}$, $\overline{n^2} = \bar{n}(1+\bar{n})$ and the two first cumulants then become:

$$\kappa_{[1]} = N_{DNA}\,\varepsilon_A\,\bar{n} \tag{9a}$$

$$\kappa_{[2]} = \gamma_2 N_{DNA}\,\varepsilon_A^2\,\bar{n}(\bar{n}+1) \tag{9b}$$

It is now possible to relate the effective number of molecules $N_{eff}$ and the effective brightness $\varepsilon_{eff}$, which can also be obtained from FCS measurements, to the average number $\bar{n}$ of fluorescent labels per molecule.

$$N_{eff} = \gamma_2 \frac{\kappa_{[1]}^2}{\kappa_{[2]}} = N_{DNA}\,\frac{\bar{n}}{\bar{n}+1} \tag{10a}$$

$$\varepsilon_{eff} = \frac{1}{\gamma_2} \frac{\kappa_{[2]}}{\kappa_{[1]}} = \varepsilon_A (\bar{n} + 1) \qquad (10b)$$

We point out the fact that $N_{eff}$ is smaller than the total number of molecules, $N_{DNA}$, because molecules that do not carry any fluorescent label are not observed. Consistently, $\varepsilon_{eff}$ does not tend towards zero when $\bar{n}$ does, but to $\varepsilon_A$, because the cDNA strands that remain fluorescent bear, at least, one label. The above expressions will prove useful in section 4.4 to extract information about the average number of fluorescent labels from FCS-photobleaching measurements.

### 3.2.4 Brightness distribution and higher order cumulants

Let us now consider the consequence of the brightness distribution for the factorial cumulants of higher order (3 to 5). In particular, it is interesting to find out what information the analysis of higher order cumulants could afford on the statistical distribution of brightness. We still assume that the brightness distribution results only from that of the number, $n$, of Alexa labels in cDNA strands (*i.e.* $\overline{\varepsilon^r} = \varepsilon_A^r \overline{n^r}$). Hence we have, according to eq 6:

$$\kappa_{[r]} = \gamma_r N_{DNA} \varepsilon_A^r \overline{n^r} \qquad (11)$$

where $\overline{n^r}$ are the different moments of the number of label distribution. We then consider the following ratio of cumulants:

$$\frac{\kappa_{[r]}^2}{\kappa_{[r-1]}\kappa_{[r+1]}} = \frac{\gamma_r^2}{\gamma_{r-1}\gamma_{r+1}} \frac{\left(\overline{n^r}\right)^2}{\left(\overline{n^{r-1}}\right)\left(\overline{n^{r+1}}\right)} \qquad (12)$$

The first term on the right hand side of eq 12, let us call it $A(r)$, depends upon the observation volume. The second term on the right hand side is equal to 1 for a homogeneous dye solution, since in this case $n \equiv 1$. We can thus measure the first geometrical term $A(r)$ with a solution of Alexa and use it to obtain the normalized ratios of factorial cumulants, $K_{rr,r-1,r+1}$, that depends only on the moments of the number of label distribution:

$$K_{rr,r-1,r+1} = \frac{1}{A(r)} \frac{\kappa_{[r]}^2}{\kappa_{[r-1]}\kappa_{[r+1]}} = \frac{\left(\overline{n^r}\right)^2}{\left(\overline{n^{r-1}}\right)\left(\overline{n^{r+1}}\right)} \qquad (13)$$

If the distribution is Poissonian, the moments are all related to the first moment, $\overline{n}$, according to:

$$\overline{n^2} = \overline{n}(1+\overline{n}) \qquad (14a)$$

$$\overline{n^3} = \overline{n}\left(1 + 3\overline{n} + \overline{n}^2\right) \qquad (14b)$$

$$\overline{n^4} = \overline{n}\left(1 + 7\overline{n} + 6\overline{n}^2 + \overline{n}^3\right) \qquad (14c)$$

$$\overline{n^5} = \overline{n}\left(1 + 15\overline{n} + 25\overline{n}^2 + 10\overline{n}^3 + \overline{n}^4\right) \qquad (14d)$$

In this case, the coefficients $K_{rr,r-1,r+1}$ depend only on the average number of fluorophores per cDNA strand, $\bar{n}$, as can be seen in Figure 2, for $n = 3$ and 4.

Note that the upper limit of $K_{rr,r-1,r+1}$ is 1, which corresponds to a single valued distribution (n ≡ 1). It is asymptotically attained when $\bar{n}$ tends towards infinity, because in this case the relative width of the Poissonian distribution tends to zero. The upper limit of $K_{rr,r-1,r+1}$ is also attained when $\bar{n}$ tends to zero, since, in this case, the only molecules that are bright are those having 1 Alexa label and no more.

We present in section 4.3 the experimental estimation of these factorial cumulants and to what extent they can provide information on the statistical distribution of fluorescent labels.

## 4. Results and Discussions

### 4.1 Fluorescence Correlation Spectroscopy

By fitting the autocorrelation function to the appropriate model for free diffusion, the effective number of molecules in the observation volume, $N_{eff}$ and the diffusion time, $\tau_D$ of the different species (Alexa, Alexa-dCTP, short and long cDNA) can be extracted. We observed that cDNA strands diffuse about ten times slower than Alexa fluorophores (see Figure 3a and Table 1). Of course, because of their much larger molecular mass, cDNA molecules have an hydrodynamic radius much larger than Alexa fluorophores. However, a precise theoretical determination of their hydrodynamic radius should take into account their sequence and the pH, salinity of the solution, *etc* and will not be attempted here.[31] Figure 3b shows the difference between a native sample and a denatured sample of short cDNA strands from the same batch. The number of molecules from the denatured sample is almost twice that of the native sample, which confirms that short cDNA strands tend to hybridize and form dimers as mentioned in Section 2.1.

Actually, in the scope of the present paper, the most useful data we have obtained from the FCS measurement are the effective brightness of the cDNA strands, $\varepsilon_{eff}$, or, equivalently, the effective photon count rates per molecule, $CRM_{eff}$. Table 1 shows that the effective photon count rates per molecule of the long and short native strands is about twice that of free Alexa, while that of denaturated short cDNA strands is close to Alexa. The brightness of a cDNA strand bearing $n$ Alexa fluorophores is simply given by $\varepsilon_{eff} = n\varepsilon_A$, where $\varepsilon_A$ is the brightness of individual Alexa labels in cDNA (note that self-quenching between Alexa fluorophores is unlikely due to the sparse labeling obtained by our method). However, the value of $n$ cannot be deduced solely from the FCS measurements, since the brightness $\varepsilon_A$ of individual Alexa fluorophores linked to cDNA may be different from that of free Alexa molecules in solution. Nevertheless, $\varepsilon_A$ can be expressed as a function of the excited state radiative and non-radiative decay rates, because it is proportional to the fluorescence quantum yield, it:

$$\varepsilon_A \propto \eta_F = \frac{k_r^0}{k_r^0 + k_{nr}} \qquad (15)$$

Throughout this paper, we shall assume that: i) the laser intensity is sufficiently low so that the molecular transitions are not saturated and that the triplet state population is negligible; ii) the absorption cross-section from the ground state and the radiative decay rate of the excited state are the same for Alexa labels in cDNA and free Alexa dyes, hence the notation $k_r^0$. However, the non-radiative decay rate $k_{nr}$ may change when the Alexa fluorophore is attached to the cDNA backbone due to quenching mechanisms. To investigate the possible differences in photophysical properties, we performed continuous photobleaching experiments and measured the photobleaching cross-sections of our various samples.

**4.2 Continuous photobleaching**

Table 1 summarizes the photobleaching cross-section values for our four samples: Alexa dye, Alexa-dCTP, long and short cDNA strands. We see that the addition of the dCTP side group to the Alexa dye does not significantly affects its photobleaching cross-section. On the other hand, the cDNA-substituted Alexa dye exhibits a photobleaching cross-section lower by a factor of 2, compared to free Alexa.

The photobleaching cross-section $\sigma_B$ of Alexa-labeled cDNA is proportional to the photobleaching quantum yield $\eta_B$ which can be expressed as a function of the radiative decay rate $k_r^0$, the non-radiative decay rate $k_{nr}$ and the photobleaching decay rate $k_B$ of the excited state of Alexa fluorophores:

$$\sigma_B \propto \eta_B = \frac{k_B}{k_r^0 + k_{nr} + k_B} \approx \frac{k_B}{k_r^0 + k_{nr}} \qquad (16)$$

The photobleaching decay rate $k_B$ can be neglected in the denominator since it is much smaller than the two other terms. The above expression clearly shows that there are two possible reasons for the reduced photobleaching cross-section of Alexa in cDNA strands: the first one is an increase of the non-radiative relaxation rate, $k_{nr}$ (assuming a constant $k_B$ rate), due to quenching of the Alexa dye by the proximity of the cDNA backbone; the second one is a reduced photobleaching rate, $k_B$ (assuming a constant $k_{nr}$ rate), either because of a more rigid conformation or because Alexa is shielded from reactive molecules by the cDNA strand.

These two possibilities lead to different estimations of the number, $n$, of Alexa fluorophores in cDNA strands. To make these estimations, we consider the normalized photobleaching cross-sections $\alpha_B$ and the normalized brightness $\alpha_\varepsilon$ (see Table 1) obtained by dividing the values relative to cDNA to those for free Alexa molecules. These normalized quantities can be respectively written as:

$$\alpha_B = \frac{\sigma_B}{\sigma_B^0} = \frac{k_r^0 + k_{nr}^0}{k_r^0 + k_{nr}} \times \frac{k_B}{k_B^0} \qquad (17a)$$

and

$$\alpha_\varepsilon = \frac{\varepsilon_{eff}}{\varepsilon_A^0} = \frac{n\varepsilon_A}{\varepsilon_A^0} = n\frac{k_r^0 + k_{nr}^0}{k_r^0 + k_{nr}} \qquad (17b)$$

so that their ratio is given by:

$$\frac{\alpha_\varepsilon}{\alpha_B} = n \times \frac{k_B^0}{k_B} \qquad (18)$$

Let us estimate $n$ for the two possible hypotheses mentioned earlier. If we assume the photobleaching decay rate to be the same than for free Alexa dye (*i.e.* $k_B = k_B^0$), the ratio $\alpha_\varepsilon / \alpha_B$ directly gives the number, $n$, of Alexa labels on the strands, so that we deduce, from Table 1, $n \cong 4$. Conversely, in the case of a constant non radiative decay rate (*i.e.* $k_{nr} = k_{nr}^0$), the normalized brightness, $\alpha_\varepsilon$, directly gives the number of Alexa labels, that is $n \cong 2$. Note that these are extremes values, since the truth may be that both $k_B$ and $k_{nr}$ differ between free and Alexa fluorophores in cDNA. When comparing theses values of *n* to the number of C bases (74 for the long strands and 2 times 45 for the native short strands), we find that the degree of labeling is probably quite low, of the order a few %, in agreement with other reports[5, 32, 33].

Finally, from the FCS and continuous photobleaching results, one cannot conclude precisely about the number of Alexa fluorophores labeling the cDNA strands. This is because information is lacking about the brightness of the Alexa molecules linked to cDNA. It is also worthwhile to note that the number of Alexa labels is likely to be statistically distributed and not single valued as implicitly assumed so far.

**4.3 Statistical distribution of the number of fluorescent labels**

The statistical properties of the molecular brightness $\varepsilon$ reflects the distribution of the number *n* of labeled C-bases, which in turn results from the statistical nature of the incorporation of the labeled C bases during the cDNA polymerization. So far, different tools have been developed to study the molecular brightness through the intensity fluctuations.[17-20, 22] Among this family of methods, FCA has the advantage, compared to PCH or FIDA, to provide analytical expressions of the moments (more precisely, the factorial cumulants) of the photon count distribution *versus* the moments of the brightness distribution (see eq 6). This is especially convenient in the present situation, since the cumulants are directly related to the moments of the statistical distribution of the number of labels (eq 11). Assuming *n* to be distributed according to a Poisson law, which is justified by the weak degree of labeling, it seems straightforward to deduce the mean number of labels per strand, $\bar{n}$, from the cumulants, by inserting eq 14 in eq 11. In practice, the accuracy of the cumulants that one can compute from the experimental data is limited by the photon count rate and the stability of the samples (that prevents long-term acquisition), so that it is not be possible to determine $\bar{n}$ precisely. Nevertheless, the cumulant analysis is still interesting to establish the fact that the brightness of the cDNA molecules is statistically distributed and that the corresponding distribution is consistent with a Poisson one with a small value of $\bar{n}$. For this purpose, we have used the normalized ratios of factorial cumulants, $K_{rr,r-1,r+1}$, introduced in the theoretical Section 3.2.4 (see eq 13) and compared the theoretical values, to the experimental ones (Figure 2).

Due to the uncertainty of the experimental results, that increases with $r$, the meaningful ratios $K_{rr,r-1,r+1}$ are limited to $r = 3$ and 4. We did not consider $r = 2$, because the ratio $K_{22,1,3}$ is always found lower than the values expected for a Poissonian distribution. This is attributed to the fact that $\kappa_{[1]}$ (at the denominator of $K_{22,1,3}$) may be increased by scattered light and other parasitical photons, while higher order cumulants are not affected by photons following a Poissonian distribution (which is the case of this unwanted light).

As an experimental result, we found $K_{33,2,4} = 0.88 \pm 0.08$ and $K_{44,3,5} = 0.9 \pm 0.1$. These values are to be compared to the theoretical curves, as shown in Figure 2. Despite significant uncertainty and variability, our results shows that the cDNA strands are not uniformly labeled and that $\bar{n}$ is very probably limited to a small value. This is an important conclusion that only a detailed intensity fluctuation analysis make it possible to do. However, the experimental data are such that it is not possible to conclude about the mean number of labels, so that an additional technique is needed. This is the motivation of the sequential photobleaching experiments.

## 4.4 Sequential photobleaching experiments

### 4.4.1 cDNA labeling

Up to now, some uncertainties remain concerning the estimation of the number of Alexa fluorophores per cDNA strand, that depends upon the assumption of a change in the radiationless decay rate ($k_{nr}$) or in the photobleaching rate ($k_B$), when Alexa fluorophores are linked to cDNA. Hereafter, we provide an estimation that just relies on the hypothesis that the number of fluorophores is statistically distributed according to a Poisson law. To this avail, we performed FCS measurements on a small volume of sample which is sequencially photobleached. In this way, we measured the effective number of molecules $N_{eff}$ and the count rate per molecule $CRM_{eff}$ at different values of the total count rate $CR$. The total count rate is a control parameter here, since it can be made to decrease using photobleaching.

To fit the experimental results, we use the expressions derived in the theoretical part (Section 3.2.3) that relate the effective number of molecules $N_{eff}$ and the effective brightness $\varepsilon_{eff}$ obtained via FCS to the average number $\bar{n}$ of fluorescent labels (see eqs 10a and 10b). Note that $\bar{n}$ refers to the number of labels at a specific step of the photobleaching process, so that it varies during the experiment, as opposed to the number of labels initially carried by the cDNA strands which is characteristic of the sample. We need to introduce the total count rate $CR$ in eq 10, since it is the quantity we directly control. $CR$ can be written using the number of cDNA strand $N_{DNA}$ and the count rate per Alexa fluorophore $CRM_A$,

$$CR = N_{DNA}\,\bar{n}\,CRM_A = N_{DNA}\,\bar{n}\,\varepsilon_A / \delta t \qquad (19)$$

where the values of the brightness and the count rate per molecule are related by the binning time, $\delta t$. Now $\bar{n}$ can be eliminated from eq 10, so that the effective number of molecules, $N_{eff}$ and count rate per molecule, $CRM_{eff}$ (i.e. $\varepsilon_{eff}/\delta t$) are expressed as a function of the total count rate, $CR$:

$$N_{eff} = N_{DNA}\,\frac{CR}{N_{DNA}\,CRM_A + CR} \qquad (20a)$$

$$CRM_{eff} = CRM_A \left(1 + \frac{CR}{N_{DNA} \, CRM_A}\right) \quad (20b)$$

Note that these two eqs are in fact strictly equivalent to each other, since $CRM_{eff} = CR/N_{eff}$. The parameters $N_{DNA}$ and $CRM_A$ can be obtained by fitting the experimental curves, $CRM_{eff}(CR)$ and $N_{eff}(CR)$. Then we finally obtain the average number of labels initially present before photobleaching (*i.e.* when the total count rate is still $CR_0$):

$$\bar{n} = \frac{CR_0}{CRM_A \, N_{DNA}} \quad (21)$$

The experimental data are shown in Figure 4. The effective count rate per molecule, $CRM_{eff}$, decreases linearly as a function of the total count rate. According to eq 20b, we obtain, from the slope and the value at $CR = 0$, $N_{DNA}$, the effective number of cDNA molecules (fluorescent or not) inside the confocal volume and $CRM_A$, the count rate per Alexa fluorophore. Then, using these values, the theoretical curve for $N_{eff}$ given by eq 20a is drawn. The experimental behavior of $N_{eff}(CR)$ is shown to be perfectly predicted by this equation, with no variable parameter. Finally, using eq 21, we can deduce from these fitting parameters the mean value of the Poisson law describing the distribution of number of fluorophores initially present on the cDNA strands: $\bar{n} = 1.80 \pm 0.22$ for the long cDNA strand (Figure 4a is a typical example) and $\bar{n} = 1.49 \pm 0.46$ for the short cDNAs. We also find $\bar{n} = 0.49$ for the denaturated short cDNAs (see an example in Figure 4b). Note that these low values of $\bar{n}$ can explain some of the observations made in FCS: the effective brightness of the denaturated strands is not exactly half that of the native strands, while their effective number of molecules is not exactly twice that of the native strands (see section 4.1). According to eq 10, when $\bar{n}$ is divided per 2, the effective brightness is decreased by a factor smaller than 2, while the effective number of molecules is lowered compared to the actual number of molecules. It is also worthwhile to note that $\bar{n}$ is a mean value calculated over the population of all the cDNA strands, *i.e.* including the dark cDNAs, having no Alexa fluorophore. Therefore, the visible cDNAs have a mean number of Alexa labels slightly larger, since it is given by $\bar{n}/(1-e^{-\bar{n}})$.

Let us comment on the shape of the experimental curves shown in Figure 4. The decrease of $CRM_{eff}$ versus $CR$, is more pronounced for the long strands (Figure 4a) than for the denaturated short strands which are labeled with a smaller number of fluorophores (Figure 4b). The effective number of molecules, $N_{eff}$, tends to be constant when there are many fluorophores per strand, as can be seen for long strands in Figure 4a. Then it becomes proportional to the count rate, $CR$, when only a few fluorophores are left. These observations could be understood by considering two limiting cases. The first one is that of a large number of fluorophores per cDNA strand: as we start to photobleach, the effective count rate per molecule decreases because cDNA strands lose fluorophores (eq 10b), but since no cDNA molecule becomes totally dark (because the strands are still carrying numerous fluorophores), the effective number of molecules remains almost constant (eq 10a with $\bar{n} \gg 1$). The other limiting case occurs for a population of cDNA strands labeled with such a small number of fluorophores, that the probability to find a strand with two labels is negligible. In that case, the effective count rate per molecule is almost constant and equal to that of one label (see eq 10b), while the effective number of (visible) molecules decreases (eq 10a with $\bar{n} \ll 1$).

In fact, we see from eq 10 that it is necessary to cross over the transition between the two limiting cases in order to determine the mean number of labels per strand: as a matter of fact, if $\bar{n}$ is very large, one can only determine $N_{DNA}$, while when $\bar{n}$ is very small, one can only determine $\varepsilon_A$ (*i.e.* $CRM_A$). The same conclusion can be drawn from eq 20: $N_{DNA}$ and $CRM_A$ can only be determined in the transition regime and both quantities are necessary to calculate $\bar{n}$ (see eq 21). Consequently the sequential photobleaching method is especially well adapted for situations where the mean number of fluorescent entities per diffusing species is limited (for a very large number of labels, one would have to photobleach the solution till the transition regime is reached, which can be difficult).

**4.4.2 From cDNA labeling to photophysical properties**

To conclude this section, we come back to the photophysical changes caused by the incorporation of Alexa fluorophores in cDNA. We can use the values of the average number of fluorescent labels obtained in the last paragraph to deduce the photobleaching decay rate, $k_B$ and the non-radiative decay rate, $k_{nr}$ (in Section 4.2, we were left with two indistinguishable possibilities). Using eq 10b, $\varepsilon_{eff} = (\bar{n} + 1)\varepsilon_A$, eq 17b must be written again to take into account the statistical distribution of the number of fluorophores, so that, contrarily to eq 18, we obtain:

$$\frac{k_B}{k_B^0} = \frac{\alpha_B(\bar{n}+1)}{\alpha_\varepsilon} \tag{22}$$

Using the values displayed in Table 1 for $\alpha_B$ and $\alpha_\varepsilon$ and the values of $\bar{n}$ from above, we find $k_B/k_B^0 = 0.72$ for long cDNA and 0.65 for short non-denatured cDNA. This result suggests that the photobleaching decay rate $k_B$ lowers when Alexa fluorophore is linked to cDNA, which is consistent with observations made by Widengren and Schwille.[34] According to these authors who characterized, by FCS, the cyanine dye Cy5 (of which Alexa is a derivative), the rates of intersystem crossing and triplet state decays were noticeably reduced for the DNA substituted dye. This fact may reflect that the dye is shielded by the cDNA strand from the oxygen molecules of the aqueous solution. Moreover, inserting the values obtained for $k_B/k_B^0$ in the expression of $\alpha_B$ (eq 17a), we find that the ratio of the decay rates of the excited state $(k_r^0 + k_{nr})/(k_r^0 + k_{nr}^0)$ is 1.36 for long cDNA and 1.33 for short non-denatured cDNA. Therefore, we can infer that the non-radiative decay rate of Alexa is increased when linked to cDNA, so that the brightness of individual fluorophores is reduced, that is: $\varepsilon_A = 0.74 \varepsilon_A^0$ for long cDNA and $\varepsilon_A = 0.75 \varepsilon_A^0$ for short non-denatured cDNA. Note that the decay rate of the excited state of Alexa could be directly measured by lifetime measurements (since $k_r^0 + k_{nr} = 1/\tau$ ).

## 5. Conclusion

We have introduced a theoretical framework in order to characterize multiple labeling using fluorescence data, independently of any assumption about the photophysical properties of the fluorophores (apart from the fact that we have neglected non linear processes and singlet-triplet transitions, because of the moderate power of our laser beam). The brightness of the cDNA strands could not have been straightforwardly used to scale the number of fluorescence labels for two reasons: i) one does not know the brightness of a single Alexa fluorophore in cDNA; ii) the brightness depends upon the statistical distribution of the number of labels. Thus, photobleaching was used as a control parameter to analyze the variations of the number of molecules and brightness *versus* the total count rate. The approach is somewhat analogous to that often used in PCH (or FIDA) analysis where one modifies the biological conditions or performs titration experiments to monitor the molecular brightness. In addition, like for these methods, the molecular brightness we refer to is an effective brightness representing a non-homogeneous sample. Whatever is the exact method used to analyze the fluctuations of fluorescence, the S/N does not make it possible to resolve the distribution of brightnesses.[35] Our underlying model relies on the hypothesis that the number of labels follows a Poissonian distribution, which is justified by the large number of sites available for labeling (C bases) compared to the number of sites actually labeled.[14] Such a distribution has the important property to be preserved during photobleaching. As a result, we found a mean value of 1.5 and 1.8 Alexa labels for the short (hybridized) and long cDNA strands, that corresponds to a labeling efficiency of less than 2%. The consequence of such a weak degree of labeling is that the brightness necessarily fluctuates quite a lot, together with the fact that a significant amount of strands are not labeled at all. For instance, for a mean value of 2 labels per cDNA and a brightness proportional to the number of labels, each species with 1 to 5 fluorophores contribute in a comparable way to the total fluorescence signal!

When combining our approach with independent measurements of the photobleaching cross-section and FCS, we got additional information about the interplay between the different decay rates of the excited state of the labeling dye: we thus found that, when incorporated in cDNA, the photobleaching decay rate is decreased by 30%, while the non-radiative decay rate increases such that the brightness is lowered by 25%. These photophysical outputs are usually provided by dedicated single molecule experiments,[36-38] or time resolved ones.[39-41]

The method presented in this paper can be applied to DNA samples of low concentration, while standard measurements of labeling efficiency are based on absorption spectroscopy and thus require amplification of the genetic material.[33] More generally, the quantification of multiple labeling is a very common problem for protein or DNA analysis involved in biochemistry, pharmacology, medicine, *etc.* The question is especially crucial with the development of ultra sensitive techniques to detect biomolecules, among which, single molecule measurements play a central role but are also, unfortunately, quite sensitive to multiple labeling. We believe that our approach is very general and relatively easy to implement, since it implies bulk measurements with concentrations in the nM - µM range and does not need single molecule immobilization and imaging. For instance, for protein interactions, the model could be extended to situations where the number of label distribution follows different statistical laws.[12,42]

## 6. Acknowledgements

This project was funded by the French Agence Nationale de la Recherche under contract ANR-07-GPLA-013-001.

The authors also gratefully thank Jie Gao for helpful and fine technical assistance on the FCS experiment.

**Figure captions and Table**

**Figure 1.** Time-derivative of the decay of the fluorescence signal, $F(t)$, due to photobleaching, on a semi-log scale measured with 4 mW excitation power. Results are shown for three samples: Alexa (open circles), Alexa-dCTP (open squares) and Alexa labeled long cDNA strands (open triangles). Fluorescence decays obtained from short cDNA strands are similar to that of long strands (data not shown).

**Figure 2.** Normalized ratios of factorial cumulants, $K_{33,2,4}$ (dashes) and $K_{44,3,5}$ (dots), *versus* $\bar{n}$. The continuous curves correspond to the theoretical calculations done by assuming a Poissonian distribution with a mean value of $\bar{n}$ Alexa fluorophores per cDNA strand. The vertical error bars represent the experimental values of the normalized ratios with their uncertainties.

**Figure 3** Example of experimental autocorrelation functions (symbols), acquired with different species and concentrations, with their fits for free diffusion (solid lines). (a) autocorrelation functions for Alexa (open triangles) and long cDNA strands (open circles), exemplifying the difference of diffusion times; they are normalized so that $G(0)=2$. (b) autocorrelation functions of short cDNA strands, the amplitude of which is about twice when the strands are hybridized (open circles *versus* open triangles).

**Figure 4** Typical count rate per molecule ($CRM_{eff}$, solid squares) and effective number of molecules ($N_{eff}$, open circles) as a function of the total count rate, $CR$, determined from FCS measurements during photobleaching. (a) Long cDNA strands. (b) Denatured short cDNA

strands. The count rates per molecule are fit with eq 20b and the initial number of labels per cDNA molecule ($\bar{n}$) is then determined from eq 21.

**Table 1** Photophysical parameters of Alexa, Alexa-dCTP and cDNA strands, determined by continuous photobleaching and FCS experiments.[a]

|  | $\sigma_B$ ($10^{-22}$ cm$^2$) | $CRM_{eff}$ (kHz) | $\tau_D$ (µs) | $\alpha_B$ | $\alpha_\varepsilon$ |
|---|---|---|---|---|---|
| Alexa | 10.6 ± 0.2 | 9.5 ± 0.3 | 59 ± 2 | 1 | 1 |
| Alexa-dCTP | 11.4 ± 0.2 | 10.2 ± 0.3 | 68 ± 3 | 1.07 | 1.07 |
| Long cDNA | 5.7 ± 0.1 | 19.4 ± 2 | 632 ± 60 | 0.53 | 2.05 |
| Short cDNA | 5.2 ± 0.1 | 18.0 ± 2 | 490 ± 30 | 0.49 | 1.89 |
| Denatured short cDNA |  | 11.2 | 410 |  | 1.18 |

[a] $\sigma_B$ is the photobleaching cross-section; $CRM_{eff}$ is effective the photon count rate per molecule, normalized at 1 µW of laser power; $\tau_D$ is the diffusion time; $\alpha_B$ and $\alpha_\varepsilon$ are the normalized photobleaching cross-section and brightness, that represent the photobleaching cross-section and brightness ($\varepsilon_{eff}$) of the various Alexa labeled entities (dCTP, cDNA), relative to the corresponding values for free Alexa molecules ($\sigma^0_{B,A}$ and $\varepsilon^0_A$), i.e. $\alpha_B = \sigma_B / \sigma^0_{B,A}$ and $\alpha_\varepsilon = \varepsilon_{eff} / \varepsilon^0_A$.

**Figure 1**

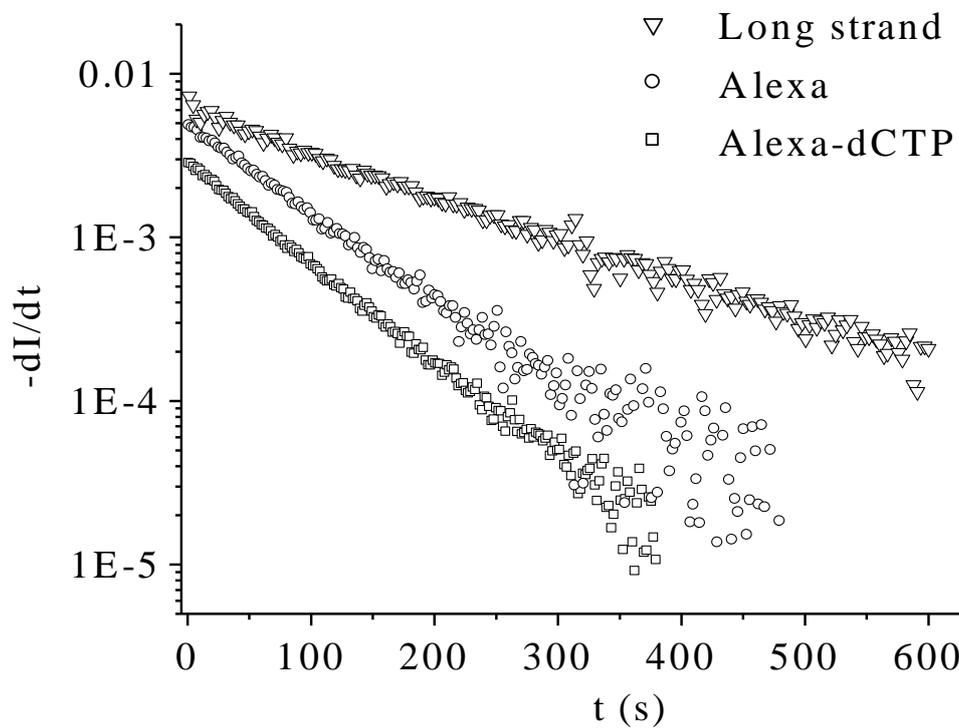

**Figure 2**

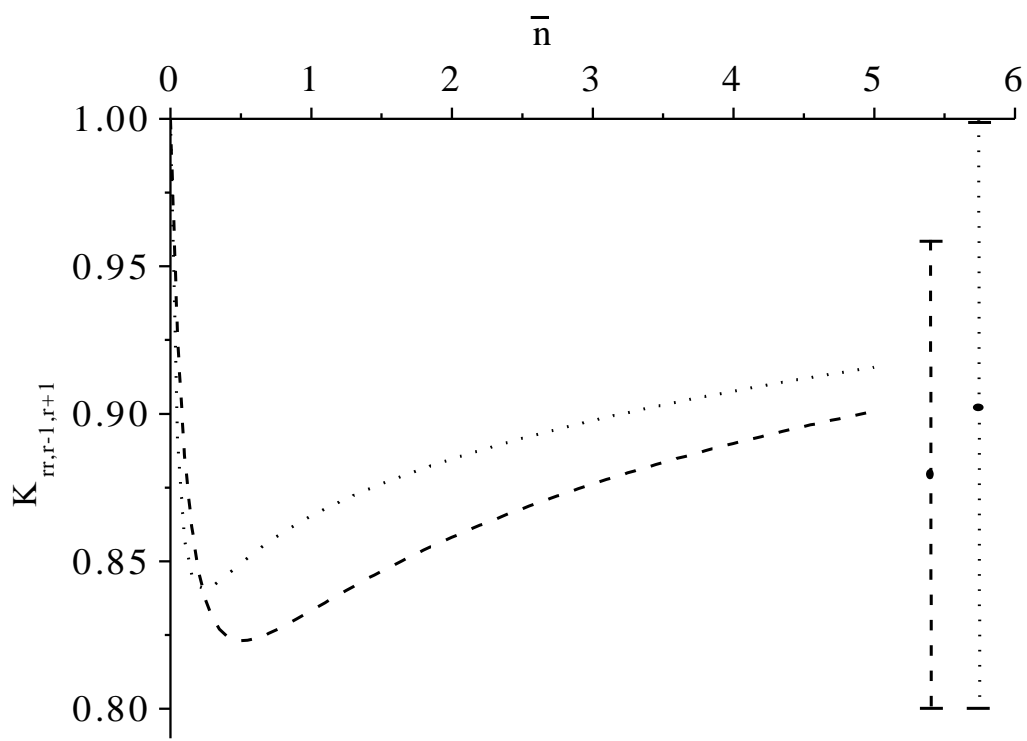

**Figure 3**

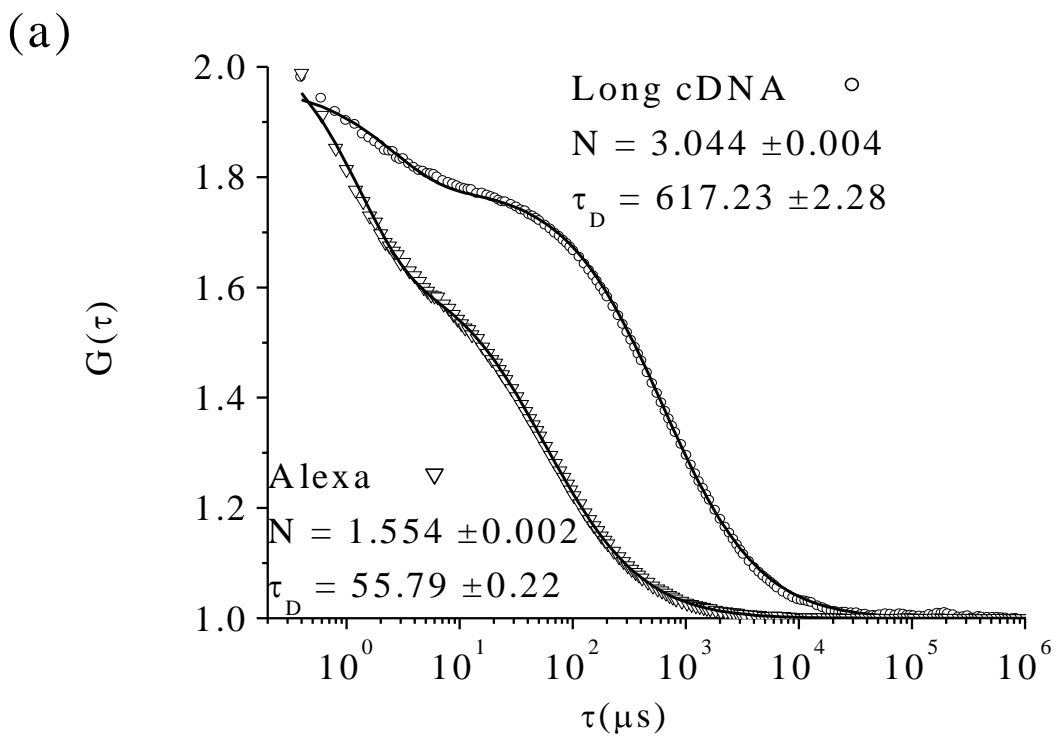

**Figure 3**

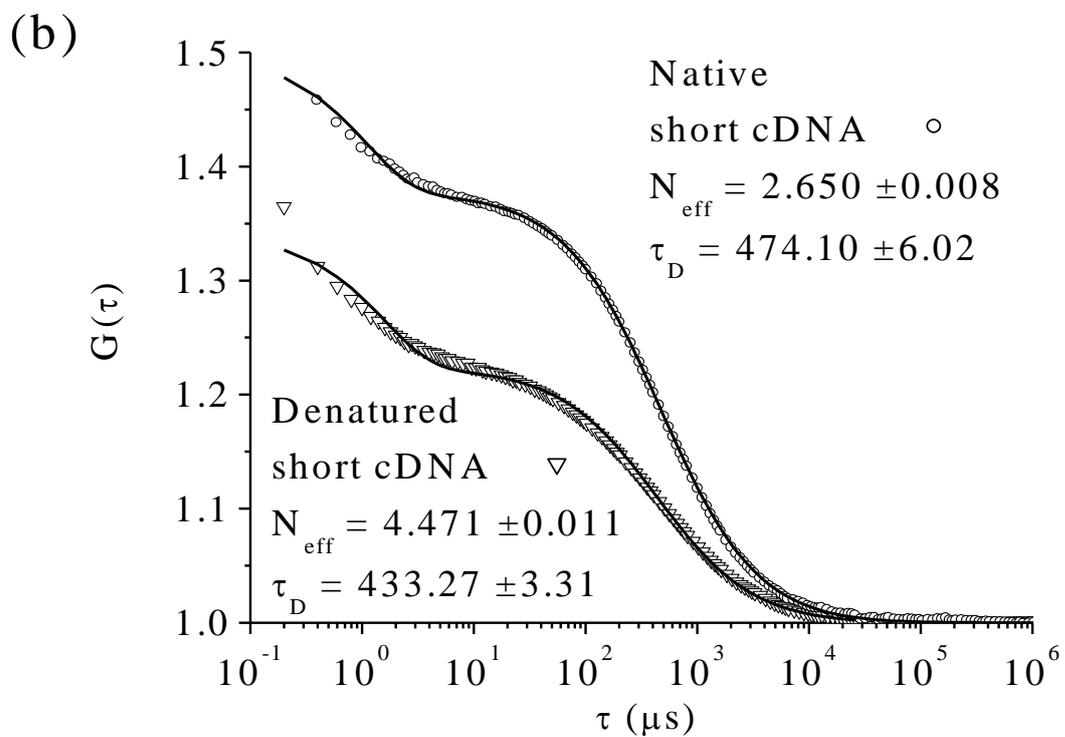

**Figure 4**

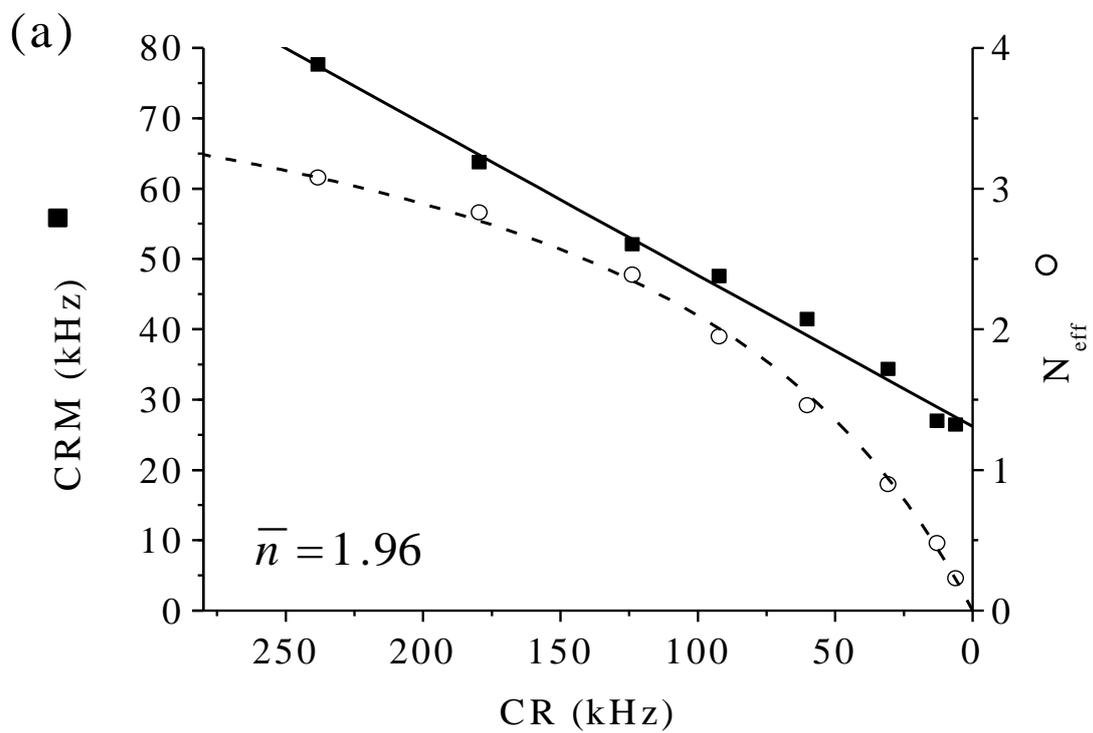

(a)

**Figure 4**

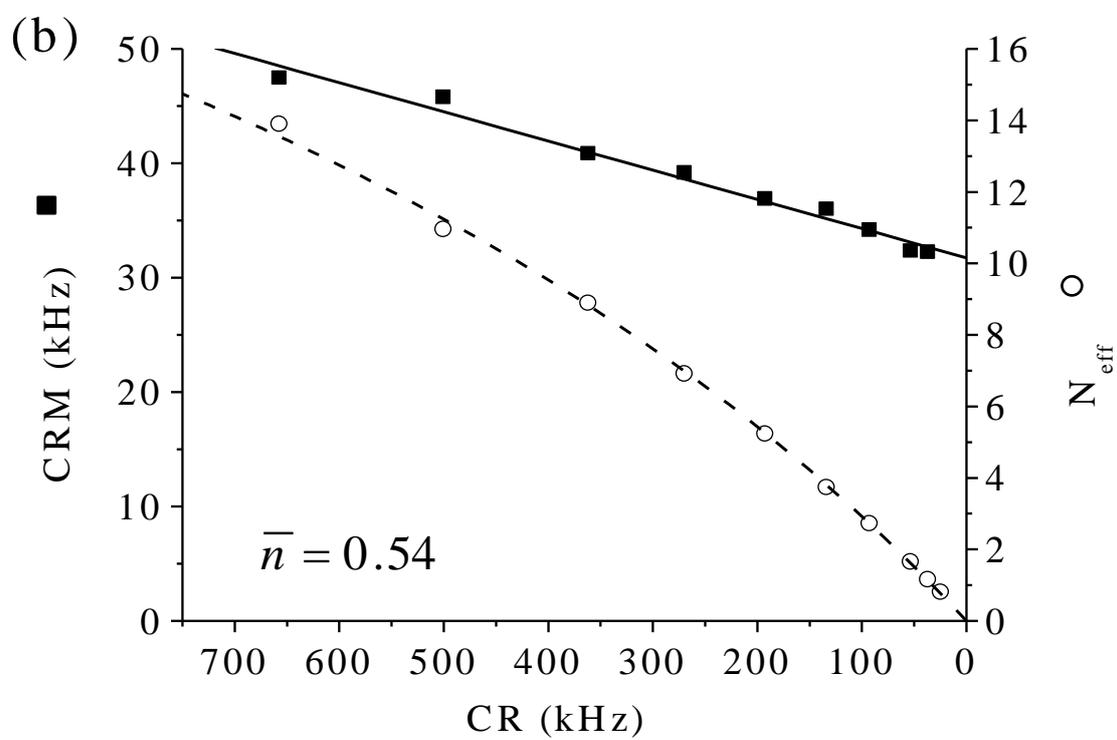

(b)

**Table of Contents**

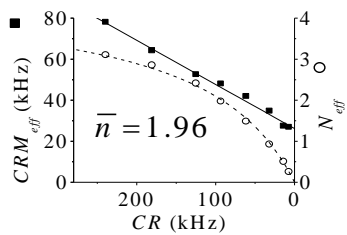

Typical count rate per molecule ($CRM_{eff}$, solid squares) and effective number of molecules ($N_{eff}$, open circles) as a function of the total count rate, $CR$, determined from FCS measurements during photobleaching of a solution of cDNA strands. These strands are 306 base long, including 74 dCTP, some of them being labeled by Alexa-Fluor-647. The fit of the effective count rate per molecule (solid line) provides the initial number of labels per cDNA molecule ($\bar{n}$). The dashed curve is the effective number of molecules that corresponds to this fit.